\documentclass{elsart}
\usepackage{graphics,natbib,graphicx,psfig,amssymb} 
\journal{New Astronomy}
\input psfig.sty
\begin{document}
\begin{frontmatter}
\title{Galactic Longitude Dependent Galactic Model Parameters}
\author[istanbul]{S. Bilir\corauthref{cor}},
\corauth[cor]{corresponding author.}
\ead{sbilir@istanbul.edu.tr}
\author[istanbul]{S. Karaali},
\author[istanbul]{S. Ak},
\author[istanbul]{E. Yaz}
\author[beykent]{and E. Hamzao\u glu}
\address[istanbul]{Istanbul University, Faculty of Sciences, Department 
of Astronomy and Space Sciences, 34119 University, Istanbul, Turkey}
\address[beykent]{Beykent University, Faculty of Engineering and Architecture, 
Department of Computer Engineering, Beykent 34500, Istanbul, Turkey\\}

\begin{abstract}
We present the Galactic model parameters for thin disc estimated by Sloan Digital Sky Survey ($SDSS$) data of 14 940 stars with apparent magnitudes $16<g_{o}\leq21$ in six intermediate latitude fields in the first Galactic quadrant. Star/galaxy separation was performed by using the $SDSS$ photometric pipeline and the isodensity contours in the $(g-r)_{0}-(r-i)_{0}$ two colour diagram. The separation of thin disc stars is carried out 
by the bimodal distribution of stars in the $(g-r)_{o}$ histogram, and the absolute magnitudes were evaluated by a procedure presented in the literature \citep{Bilir05}. Exponential density law fits better to the derived density functions for the absolute magnitude intervals $8<M(g)\leq9$ and $11<M(g)\leq12$, whereas sech/sech$^{2}$ laws are more appropriate for absolute magnitude intervals $9<M(g)\leq10$ and $10<M(g)\leq11$. We showed that the scaleheight and scalelength are Galactic longitude dependent. The average values and ranges of the scaleheight and the scalelength are $<H>=220$ pc ($196\leq H \leq 234$ pc) and $<h>=1900$ pc ($1561\leq h \leq 2280$ pc) respectively. This result would be useful to explain different numerical values claimed for those parameters obtained by different authors for the fields in different directions of the Galaxy.    
\end{abstract}

\end{frontmatter}

\section{Introduction}
The study of Galactic models and parameters have long history. \cite{BS80} fitted observations to two-component Galactic model, namely disc and halo, 
while \cite{GR83} successfully fit their observations to a Galactic model introducing a third component, i.e. the thick disc. It should be noted 
that the third component was a rediscovery of the ``Intermediate Population II" first described in the Vatican Proceedings review of \cite{O58}. Due to their importance, Galactic models have been primary concern and research topic for many research centers: As they can be used as a tool in order to understand the formation and evolution of the Galaxy.

Different research groups have been using different methods to determine Galactic model parameters (Table 1). For example, \cite{Chen01} and \cite 
{Siegel02} give 6.5-13 and 6-10 per cent, respectively, for the relative local space density of the thick disc. However, we think that an appropriate procedure would lead us to expect a smaller range and/or a unique value with a small error. 

In the previous studies \citep{KBH04, Bilir06} we estimated Galactic model parameters for stars of different Galactic populations and absolute magnitude intervals and we found that the range of these models are rather small relative to the ones appeared in the literature. It gives the indication that this procedure refines the Galactic model parameters. However such a result may not be considered in the model estimation due to the 
contamination introduced by the other populations. In the present study we estimated the Galactic model parameters only for the thin disc and our previous results are confirmed. Also, we showed (Section 5.1) that the contamination of the thick disc is rather small. Additionally we discussed the dependence of the Galactic model parameters on the Galactic longitude. For this, we used the homogeneous $SDSS$ data for stars in six intermediate Galactic latitude fields in the first Galactic quadrant. The range of the latitudes of the fields is comparatively small, $41^{o}\leq b \leq52^{o}$, whereas the longitudes of the fields lie between $5^{o}$ and $83^{o}$. Hence, any significant difference between the values of a given Galactic model parameters derived for stars in the same absolute magnitude interval for two fields could be attributed to the effect of the Galactic longitude, under the condition that our data are not contaminated by other populations, i.e. thick disc. The sample stars are at distances less than $r\sim2$ kpc relative to the Sun, and as stated above, the thick disc contamination is rather small. Thus, we can say that Galactic model parameters are longitude dependent, at least for the six fields investigated in the present study.      

In Section 2 we describe the data and reductions, in Section 3 we introduce the density laws adopted in the present study. Section 4 provides the absolute magnitude determination, and the evaluation of the density functions for thin disc of six fields. In Section 5 we estimate the Galactic model parameters, and the final conclusion is presented in Section 6.   

\section{Data and reductions}

The data were taken from $SDSS$, Data Release 3 (DR3), on the WEB\footnote{http://www.sdss.org/dr3/access/index.html} of six fields with  intermediate Galactic latitude ($41^{o}\leq b \leq52^{o}$) in the first Galactic quadrant (Table 2). $SDSS$ magnitudes $u$, $g$, $r$, $i$, and $z$ were used for totally 113 380 stars in six fields equal in size (10 deg$^{2}$) except Field F6 (20 deg$^{2}$), down to the limiting magnitude of $g_{0}=21$. The mean $E(B-V)$ colour excess for each field is less than 0.06 except for the Field F1, $E(B-V)=0.10$. The $E(B-V)$ colour excesses were individually evaluated for each sample source making use of the maps of \cite{Schlegal98} through $SDSS$ query server and this was reduced to total absorption $A_{V}$ via equation (1).

\begin{equation}
A_{V}=3.1 E(B-V).\\
\end{equation}

In order to determine total absorptions, $A_{m}$, for the $SDSS$ bands, we used $A_{m}/A_{V}$ data given by \cite{Fan99}, i.e. 1.593, 1.199, 0.858, 0.639 and 0.459 for $m$=$u$, $g$, $r$, $i$ and $z$, respectively. Thus, the de-reddened magnitudes, with subscript 0, are 

\begin{equation}
u_{0}=u-A_{u},~~g_{0}=g-A_{g},~~r_{0}=r-A_{r},~~i_{0}=i-A_{i},~~z_{0}=z-A_{z}.
\end{equation}
All the colours and magnitudes mentioned hereafter will be de-reddened ones.

According to \cite{Chen01}, the distribution of stars in the apparent magnitude-colour diagram, $g_{o}-(g-r)_{o}$, can be classified as follows. The 
blue stars in the range $15<g_{0}<18$ are dominated by thick-disc stars with turn-off at $(g-r)_{0} \sim 0.33$, and for $g_{0}>18$ the Galactic halo stars, with turn-off at $(g-r)_{0} \sim 0.2$, become significant. Red stars, $(g-r)_{0} \sim 1.3$, are dominated by thin disc stars at all apparent magnitudes. 

However, the apparent magnitude--colour diagram and the three two-colour diagrams for all objects (due shortage of space large amount of data not presented here) indicate that the stellar distributions are contaminated by extragalactic objects as claimed by \cite{Chen01}. Distinction between star/galaxy was obtained using command probPSF$_{mag}$ given in DR3 WEB page. There 1 or 0 is designated for the probability of object being a star or galaxy. Needless to say, separation of 1 or 0 strongly depends on seeing and sky brightness. Apart from the above mentioned work, simulations of \cite{Fan99} were carefully adopted in order to remove the extragalactic objects in our field. This is similar to the procedure of \cite{Juric06} who defined locus points and drew several isodensity contours in the $(g-r)_{0}$-$(r-i)_{0}$ two colour diagrams and rejected stars on the contours at distances larger than 0.3 mag relative to the corresponding locus (Fig. 1).

The colour-magnitude diagram $g_{o}$-$(g-r)_{o}$ for the final sample (stars) are given in Fig. 2. The limiting magnitude for the survey stars in Fig. 2 can be read off as $g_{0}=22$. However, the colour and magnitude errors for fainter magnitudes are larger, i.e. ($\sigma_{u-g}=0.26$, $\sigma_{g-r}=0.04$, $\sigma_{r-i}=0.04$, $\sigma_{i-z}=0.07$) for $g_{0}=21$, and ($\sigma_{u-g}=0.75$, $\sigma_{g-r}=0.06$, $\sigma_{r-i}=0.05$, $\sigma_{i-z}=0.08$) for $g_{0}=22$ hence we adopted $g_{0}=21$ as the limiting magnitude in our work for the evaluation of more reliable Galactic model parameters.

\section{Density laws}

Disc structures are usually parametrized in cylindrical coordinates by radial and vertical exponentials,
\begin{equation}
D_{i}(x,z)=n_{i}~exp(-z/H_{i})~exp(-(x-R_{0})/h_{i}),\\
\end{equation}
where $z$ is the distance from the Galactic plane, $x$ is the planar distance from the Galactic centre, $R_{0}$ is the solar distance from the Galactic centre (8.6 kpc, \cite{Buser98}), $H_{i}$ and $h_{i}$ are the scaleheight and scalelength, respectively, and $n_{i}$ is the normalized density at the solar radius. The suffix $i$ takes the values 1 and 2 as long as the thin and thick discs are considered. A similar form uses the sech$^{2}$ or sech function to parameterize the vertical distribution of the thin disc. As the secans hyperbolicus is the sum of the two exponentials, the scaleheight corresponding to the sech$^{2}$ and sech has to be multiplied by 1.08504 and 1.65745, respectively (see Appendix A and \cite{Phleps00}), for its reduction to the equivalent exponential appendix. 

Here, we compare the density laws and the derived densities only for the thin disc. Hence, only the corresponding laws and parameters are considered.

\section{Absolute magnitudes, distances and density functions for late-type 
thin--disc stars}

Contrary to the thick disc and halo stars which overlap in the apparent magnitude--colour diagram, the position of the late-type thin disc stars is very conspicuous in the diagram. In fact, the $(g-r)_{0}$ histogram for all stars in six fields show a bimodal distribution, one at $(g-r)_{0}\sim0.45$ and another one at $(g-r)_{0}\sim1.35$ (Fig. 3). The Gaussian fit shows that $(g-r)_{0}=1.10$ can be adopted as the border separating the late-type thin disc stars from the thick disc and halo couple. Thus, totally 14 940 stars with $(g-r)_{0}\geq1.10$ included into our programme as thin-disc stars (details are given in Section 5.1). 

The absolute magnitudes of the programme stars were evaluated using the following calibration \citep{Bilir05}

\begin{equation}
M(g)=5.791(\pm0.023)(g-r)_{0}+1.242(\pm0.012)(r-i)_{0}+1.412(\pm0.021)(\sigma=0.05).
\end{equation}

Combination of the absolute magnitude $M(g)$ and the apparent magnitude $g_{0}$ of a star gives its distance $r$ relative to the Sun, i.e.,

\begin{equation}
[g-M(g)]_{0}=5\log r-5.
\end{equation}

For the six fields, the logarithmic space density functions were evaluated for different absolute magnitude intervals however, only the one for the Field F4 is given in Table 3 (due to shortage of space), but all of them are shown in Fig. 4 as $D^{*}=\log D+10$, where $D=N/\Delta V_{1,2}$; 
$\Delta V_{1,2}=(\pi/180)^{2}(A/3)(r_{2}^{3}-r_{1}^{3})$; $A$ denotes the size of the field (deg$^{2}$) in question; $r_{1}$ and $r_{2}$ denote the lower and upper limiting distance of the volume $\Delta V_{1,2}$; $N$ is the number of stars; $r^{*}=[(r^{3}_{1}+r^{3}_{2})/2]^{1/3}$ is the centroid distance of the volume $\Delta V_{1,2}$; and $z^{*}=r^{*}\sin (b)$, $b$ being the Galactic latitude of the field center in question. The thick horizontal lines in Table 3 correspond to the limiting distance of completeness, $z_{l}$, which are calculated from the following equations:

\begin{equation}
[g-M(g)]_{0}=5 \log r_{l}-5,\\
\end{equation}

\begin{equation}
z_{l}=r_{l}\sin(b),\\
\end{equation}
where $g_{0}$ is the limiting apparent magnitude (16 and 21, for the bright and faint stars, respectively), $r_{l}$ is the limiting distance of completeness, and $M(g)$ is the corresponding absolute magnitude in the interval ($M_{1}$, $M_{2}$).	

\section{Galactic longitude dependent Galactic model parameters}

\subsection {Estimation of Galactic model parameters by comparison of the observation-based density functions with different density laws}

A $\chi^{2}$ method was employed to fit the analytical density laws given in Section 3 for thin disc to the observation-based space densities with the additional constraint of producing local densities consistent with those derived from Hipparcos \citep{Jahreiss97}. This procedure was applied in our previous papers (KBH, BKG). 

One must be certain that our space density functions should belong to stars of the thin disc before discussing the Galactic model parameters, evaluated by the comparison mentioned above. Our strong argument is about the distance of stars relative to the Sun: All stars are at distances less that $r \sim 2$ kpc, and most of them at distances shorter than 1.5 kpc (Fig. 4). 25 stars with distances $2<r\leq3.5$ kpc relative to the Sun (Table 3) are beyond the limiting distance of completeness which are not included into the statistics. However, fitting the observed data to the Galactic model for the field F4, as example, in Fig. 5 shows that there is a slight excess in the observational based number of stars in the distances near to the Galactic plane $1<z\leq1.5$ kpc. This means some contamination of the thick disc. Comparison of the observational based data for the field F4 with the Galactic model of \cite{Chen01} for two extreme scaleheights and solar normalizations of the thick disc, i.e. $H=600$ pc; $n_{2}/n_{1}=12$ per cent, and $H=1000$ pc; $n_{2}/n_{1}=3$ per cent (Figs. 6 and 7) confirms this suggestion. Now, the question is to find out  the amount of this contamination. The locus of stars estimated in our work for absolute magnitude intervals $8<M(g)\leq9$ and $11<M(g)\leq12$ are below the model curve of thin disc of \cite{Chen01}. Whereas for the intervals $9<M(g)\leq10$ and $10<M(g)\leq11$ there is an excess in number of stars for the distance intervals $z<500$ pc relative to the model curve thin disc, indicating some contamination of the thick disc in our work. We reduced the excess number of observed stars in Fig. 6 (Figs. 6 and 7 are rather similar) such as to fit to the predicted ones for the thin disc in the Galactic model of \cite{Chen01}, and we re-estimated the Galactic model parameters for the field F4. The contamination of the thick disc in number of stars are 7, 14, and 16 per cent for the absolute magnitude intervals $8<M(g)\leq9$, $9<M(g)\leq10$, and $10<M(g)\leq11$, respectively. The differences between three sets of parameters, for the previous Galactic model and the one with re-estimated parameters are as follows: $\Delta n^{*}=(0.01, 0.00, 0.01)$, $\Delta H=(3, 5, 1)$ pc, $\Delta h=(60,563,27)$ pc for the absolute magnitude intervals in the order given above. These small amounts show that the contamination of the thick disc is rather small. The Galactic model with the re-estimated parameters mentioned above fit to the observational based data (Fig. 8). Hence our results stated in the following sections can be attributed to as the properties of the thin disc. 

It turned out that the exponential law fitted better for the brightest and faintest intervals, i.e. $8<M(g)\leq9$ and $11<M(g)\leq12$, whereas sech and sech$^2$ laws are appropriate for the intermediate intervals, i.e. $9<M(g)\leq10$ and $10<M(g)\leq11$, respectively. The numerical values for $\chi^2_{min}$ given in Table 4, of the columns 7 and 12 substantiate our suggestion. 

The resulting Galactic model parameters are summarized in Table 4. The scaleheights for the sech and sech$^{2}$ laws are reduced in accordance with the exponential law by multiplying them with 1.65745 and 1.08504, respectively. Hence, the symbol $H$ in Table 4 corresponds to the exponential law. It is seen that the parameters are absolute magnitude-dependent for six fields. For example, the scaleheight is shorter for the brightest and faintest intervals with respect to one for the intermediate intervals. Also, the scalelength changes from one absolute magnitude interval to the other, without following any rule or pattern.

Here we present a question: Is it possible to claim that the scales substantially change with Galactic longitude? The answer of this question can be obtained by fitting one disc model to the stars in a given absolute magnitude range for six fields at the same time. Fig. 9 compares the derived logarithmic space densities for the absolute magnitude interval $10<M(g)\leq11$ for six fields with the thin disc model of \cite{Chen01}. The corresponding scaleheight and scalelength of the model are $H$=330 pc and $h$=2250 pc, respectively. The agreement between the derived data and the model is rather low. Additionally, the standard deviations are large and the resulting solar space densities for six fields are different than the  Hipparcos \citep{Jahreiss97} one for the same absolute magnitude interval (Table 5). The disagreement mentioned above indicates that the claim related to the scales change with Galactic longitude is valid.   

All error estimates in Table 4 were obtained by changing Galactic model parameters until an increase or decrease by 1 in $\chi^{2}$ was achieved \citep{Phleps00}. The parameters were tested against the luminosity function given in Table 4 (last column) and Fig. 10, where $\varphi^{*}(M)$ is the local space density for thin disc. The absence of the local space densities for thick disc and halo, which is limited to $\sim$10 per cent of the total local space density, did not affect the smooth agreement between the luminosity function and the derived from Hipparcos data \citep{Jahreiss97}.  

\subsection{Dependence of the scaleheight and scalelength on the galactic 
longitude}

An interesting result can be deduced from Table 4 where the dependence of the scaleheight ($H$), and scalelength ($h$), on the Galactic longitude 
($l$), for each absolute magnitude interval is seen. In fact, Table 6, Fig. 11 and Fig. 12 show that $H$ and $h$ change linearly from one interval to the other i.e.
\begin{equation}
H = a_{1}l+a_{0},\\
\end{equation}
\begin{equation}
h = b_{1}l+b_{0},\\
\end{equation}
where the coefficients $a_{i}$ and $b_{i}$ ($i$=1, 2) are given in Table 7. The correlation is higher for the scaleheight than for the scalelength and the correlation coefficient for the scaleheight decreases from the bright absolute magnitude intervals to the faint ones, contrary to the one for 
scalelength. The correlation coefficient of $H$ for the interval $11<M(g)\leq12$ is rather small, $R^{2}=0.09$, and the slope coefficient is negative, i.e. $a_{1}=-0.158$, in contradiction with the others. 

The dependence of the scaleheight and scalelength on the Galactic longitude has so far not been claimed anywhere in the literature. Hence it is a novelty.              

\section{Summary and discussion}
In the present study, we used the homogeneous $SDSS$ data for stars in six intermediate Galactic latitude fields in the first Galactic quadrant and derived Galactic model parameters for the thin disc. Star/galaxy separation was performed by utilizing the command probPSF$_{mag}$ in DR3 WEB page to provide each object's probability of being a star in each filter. As the quality of this separation strongly depends on the seeing and sky brightness, we adopted also the simulation of \cite{Fan99} in order to remove the extragalactic objects from our sample. This is similar to the procedure of \cite{Juric06} who define locus points and draw several isodensity contours in the $(g-r)_{0}-(r-i)_{0}$ two colour diagram and reject stars on the contours at distances larger than 0.3 mag relative to the corresponding locus. Thin disc stars were separated from the thick disc and halo stars by the bimodal distribution of stars in the $(g-r)_{0}$ histogram. Comparison of the observational based space density functions with the Galactic model of \cite{Chen01} revealed that there is a slight contamination, i.e. 7, 14, and 16 per cent for the absolute magnitude intervals $8<M(g)\leq9$, $9<M(g)\leq10$, and $10<M(g)\leq11$, respectively, of the thick disc which does not affect the resulting Galactic model parameters considerably. Thus, we obtained a sample of 14 940 thin disc stars with apparent magnitude $16<g_{0}\leq21$ and colour indice $(g-r)_{0}\geq1.10$. The absolute magnitudes were evaluated according to the new calibration appeared in one of our papers \citep{Bilir05}.   

The range of the Galactic model parameters estimated for different absolute magnitude intervals, i.e. $8<M(g)\leq9$, $9<M(g)\leq10$, $10<M(g)\leq11$ and $11<M(g)\leq12$, is rather small relative to the ones appeared in the literature. Hence, our previous results (KBH and BKG) are confirmed.

As in our previous works, we added the constraint of producing densities at the solar radius consistent with those derived from $Hipparcos$ \citep{Jahreiss97}, and showed that the exponential density law fits better to the density functions for the brightest and faintest intervals, $8<M(g)\leq9$ and $11<M(g)\leq12$, whereas sech and sech$^{2}$ are more suitable for intermediate intervals, i.e. $9<M(g)\leq10$ and $10<M(g)\leq11$. Thus, we confirmed the work of \cite{Phleps00} who showed that the observational based density functions for thin disc fit to two density laws, sech and exponential.  

Comparison of the scaleheights and scalelengths of six fields, for the same absolute magnitude intervals, with their longitudes indicate that these 
parameters are longitude dependent. However, the correlation for the scalelength is not as high as for the scaleheight one. Additionally the errors for the scalelength are larger than the errors for the scaleheight. Alternatively, one may argue that tendency of the scaleheight and scalelength may due to the effect of the thick disc and/or halo stars at different distances from the Galactic center at different longitudes. However, as we mentioned above (Section 5.1) the contamination of the thick disc is rather small. Hence this argument is out of question in our work. It is worthwhile to remind that our sample stars in a field are at distances less than $r\sim2$ kpc relative to the Sun, hence their distances from the Galactic center do not differ considerably from each other. 

One may think that the argument stated in the former paragraph is in contradiction with the double exponential structure of the thin disc. Whereas we think in a different way. We do not reject the double exponential structure of the thin disc. However, we argue that additional constraints should be added for the refinement of scaleheight and scalelength of the thin disc (and the thick disc). 

Finally, we compare scaleheights and scalelengths with the ones of \cite{Chen01}, \cite{Du03} and \cite{Phleps05}. These authors give larger 
scaleheights than the mean scaleheight, $<H>=221$ pc, in our work for fields with larger Galactic longitudes, i.e. \cite{Chen01}: $310<H<345$ pc, 
$80^{\circ}<l<180^{\circ}$ and $250^{\circ}<l<360^{\circ}$; \cite{Du03}: $H=320$ pc, $l=169^{\circ}.95$; and \cite{Phleps05}: $H=281/283$ pc, $85^{\circ}<l<335^{\circ}$. According to the relation between the scaleheight and the Galactic longitude given in this work we expect larger scaleheights for larger Galactic longitudes. Hence, the scaleheights presented in this work are in agreement with those appeared recently. The difference between our work and the others is that we argue and show the dependence of the scaleheight on the Galactic longitude. 

The mean scalelength in our work, $h=1900$ pc, is close to the ones adopted by \cite{Chen01}, $h=2250$ pc, and \cite{Phleps05}, $h=2000$ pc for estimation of the scaleheights in their work.

\section{Acknowledgments} 
This work was supported by the Research Fund of the University of Istanbul. Project number: BYP 706/07062005.  

\begin{appendix} 

\section{Appendix: Reduction of the secans hiperbolicus and secans hiperbolicus square scaleheights to the exponential one}

The secans hiperbolicus is the sum of two exponentials, i.e.

\begin{equation}
sech(x) = 2/[\exp(-x) + \exp(x)].\\
\end{equation}

Hence, the sech density law,

\begin{equation}
sech (-z /z_{o})=2/[\exp(-z/z_{o})+\exp (z/z_{o})].\\
\end{equation}

Here $z_{o}$ is the sech scaleheight. Now let us write the exponential density law, $\exp (-z/H)$, and equalize it to the sech density law:

\begin{equation}
\exp(-z/H)=2/[\exp(-z/z_{o})+ \exp(z/z_{o})],                                                     
\end{equation}

If we replace $z=H$, we find the equation $\exp(-1)=2/[\exp(-H/z_{o})+\exp(H/z_{o})]$ which procudes $H/z_{o}=1.65745$ or $H=1.65745z_{o}$. This relation is in agreement with given by \cite{Phleps00}. For sech$^{2}$, if we take the square of both sides in (A.2) and equalize it to the exponential density law:

\begin{equation}
\exp(-z/H)=4/[\exp(-2z/z_{o})+\exp(2z/z_{o})+2].\\
\end{equation}

Also if we replace $z=H$, we find $\exp(-1)=4/[\exp(-2H/z_{o})+\exp(2H/z_{o})+2]$. The solution of this equation produces $H=1.08504z_{o}$.

\section{Appendix: Reduction of the standard local space densities of Hipparcos to the $SDSS$ photometry}

The standard local space densities of Hipparcos \citep{Jahreiss97} were reduced to the $SDSS$ photometry, where the relation between $(B-V)$ and $M(V)$ derived for the main-sequence stars within 10 pc of the Sun by \cite{Henry99}, were used in order to transform $M(V)$ absolute magnitudes of the Yale Parallax Catalog \citep{vanaltena95} and the Hipparcos mission \citep{ESA97} to the $(B-V)$ colour indices, i.e.

\begin{equation}
\tiny
(B-V)=0.00099M(V)^{3}-0.038184M(V)^{2}+0.555204M(V)-1.242359.\\
\end{equation} 

Then, we used the equation of \cite{Bilir05} to derive the $M(g)$ absolute magnitudes of $SDSS$ photometry: 

\begin{equation}
\small
M(g)=M(V)+0.63359(B-V)-0.10813.
\end{equation}
Finally, the standard local space densities of Hipparcos \citep{Jahreiss97} for the absolute magnitudes of $M(g)$: 7.5, 8.5, 9.5, 10.5 and 11.5, were interpolated and is given in the column (13) of Table 4.
\end{appendix}

\begin{figure}
\begin{center}
\includegraphics[angle=0, width=165mm, height=96.38mm]{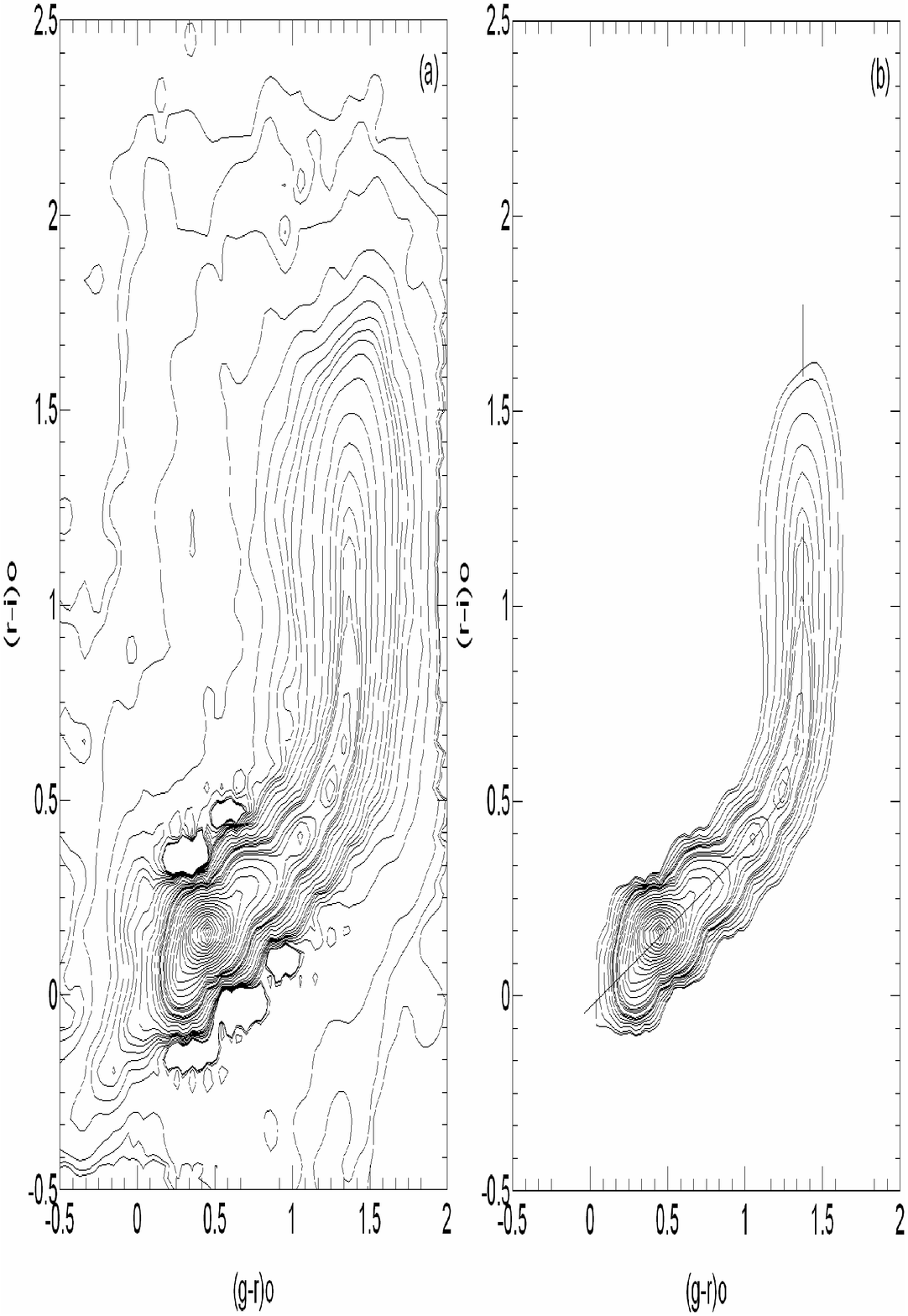}
\caption[] {Isodensity contours for the objects after performing the star/galaxy separation by using the $SDSS$ photometric pipeline (a). Objects at distances larger than 0.3 mag from the locus were rejected from the sample (b). The procedure is adopted from \cite{Juric06}.} 
\end{center}
\end{figure}
\clearpage

\begin{figure}
\begin{center}
\includegraphics[angle=0, width=150mm, height=220mm]{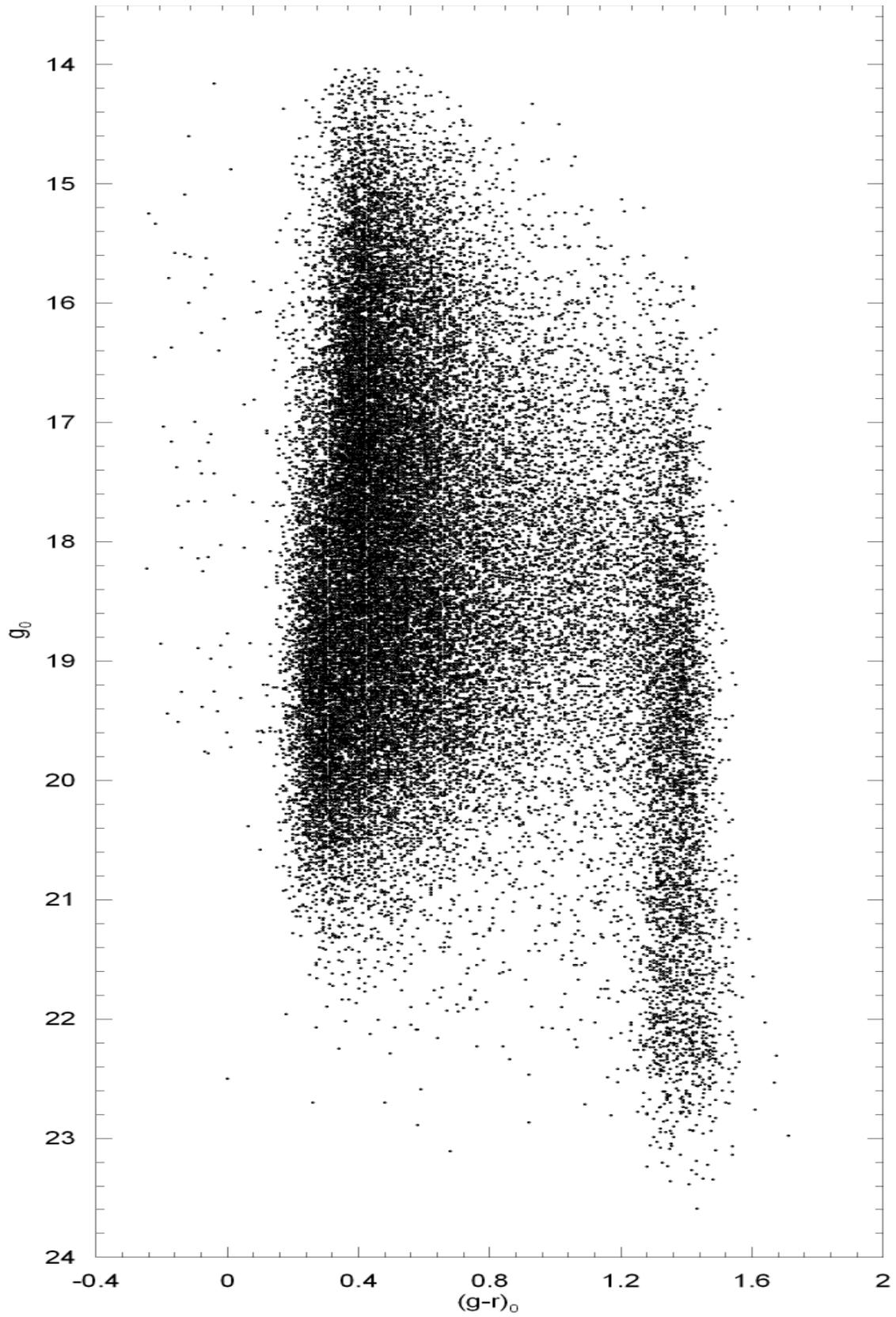}
\caption[] {Apparent magnitude colour diagram for the star sample in 
consideration.} 
\end{center}
\end {figure}

\begin{figure}
\begin{center}
\includegraphics[angle=0, width=150mm, height=94.6mm]{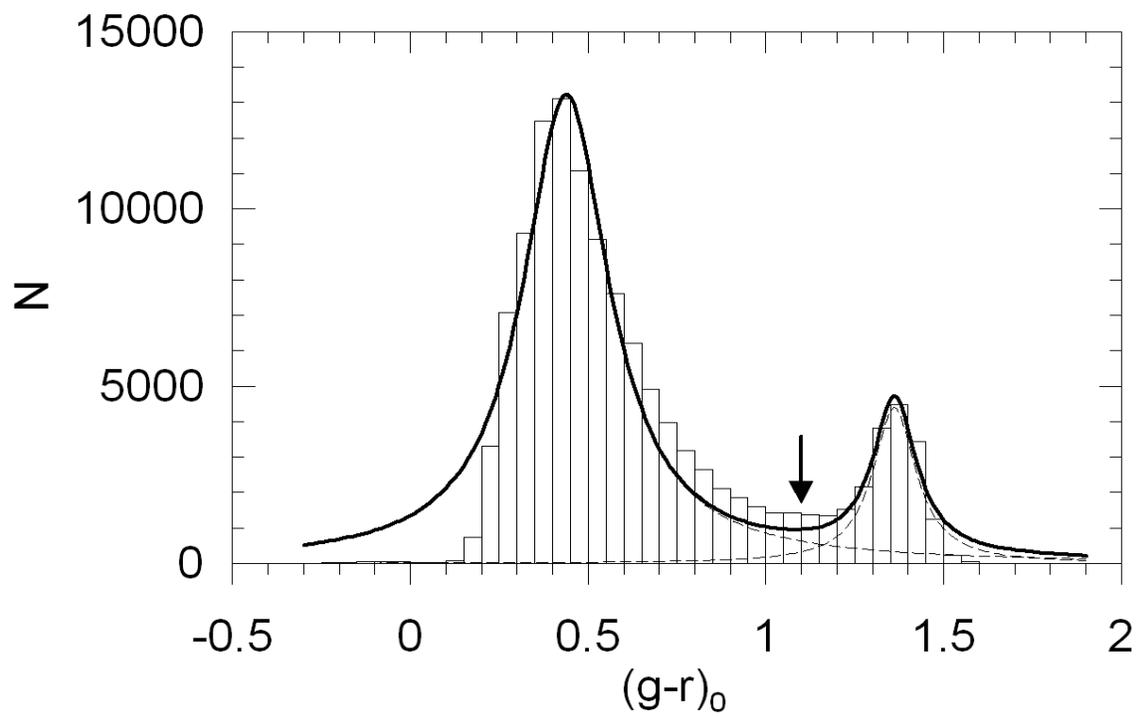}
\caption[] {$(g-r)_{o}$ histogram for the star sample used to separate the 
thin-disc stars.}
\end{center}
\end {figure}

\begin{figure}
\begin{center}
\includegraphics[angle=0, width=130mm, height=174.4mm]{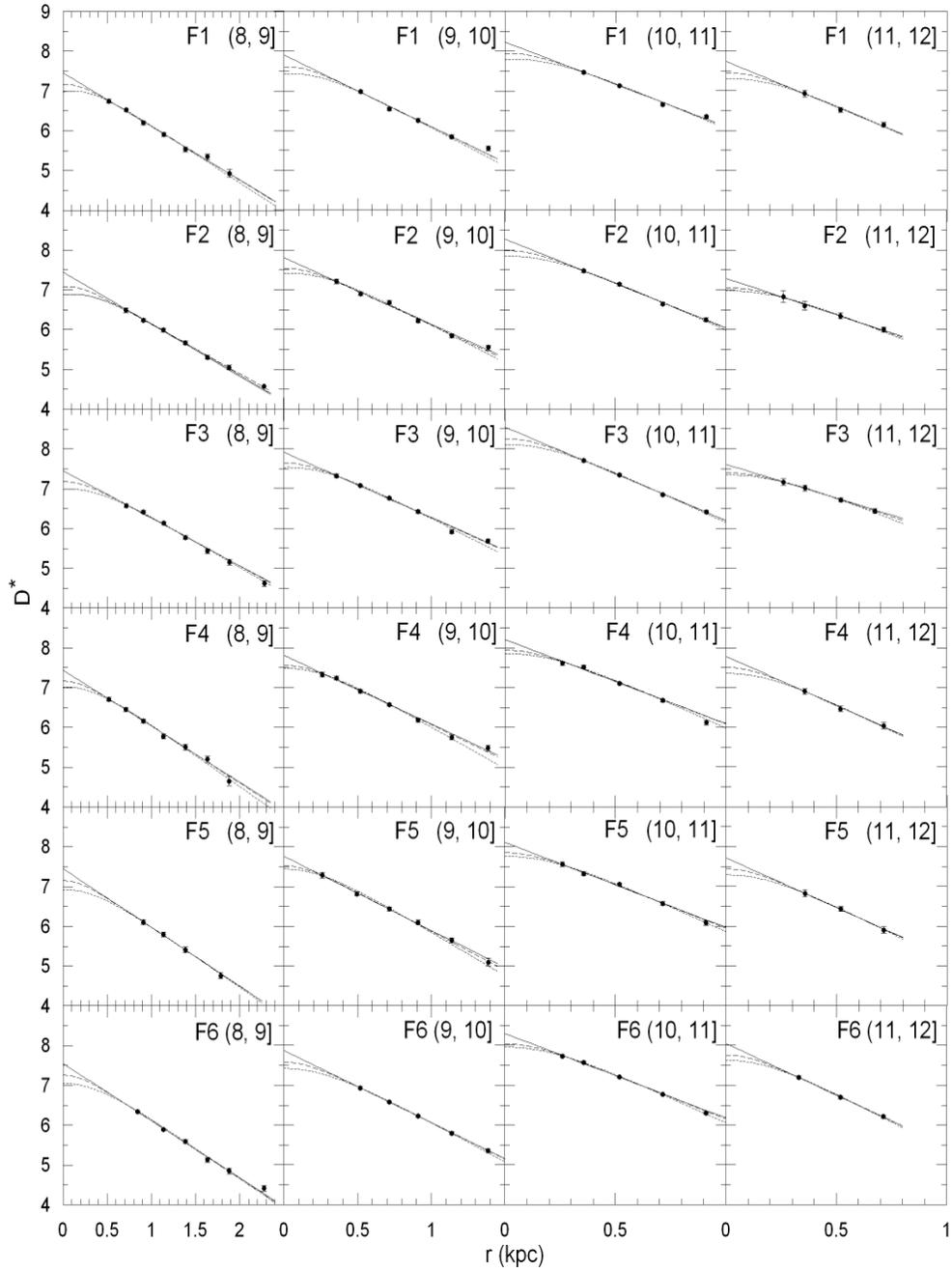}
\caption[] {Comparison of derived thin disc space density functions (symbols) 
with the best fit analytical density laws (lines) for different absolute 
magnitude intervals for six fields. The continuous curve represents the 
exponential law, the dashed curve represent the sech law and the dot-dashed 
curve represents the sech$^{2}$ law in the vertical direction.}
\end{center}
\end {figure}

\begin{figure}
\begin{center}
\includegraphics[angle=0, width=130.5mm, height=142.5mm]{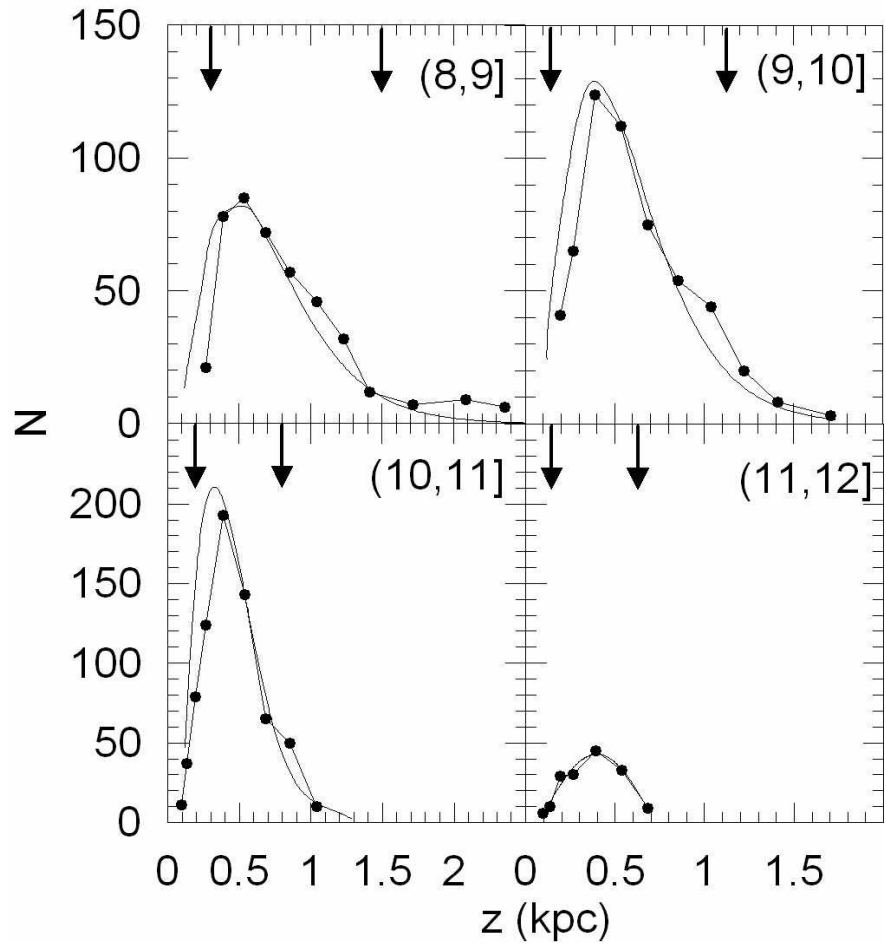}
\caption[] {Fitting of the distribution of $z$-distances for stars in four 
absolute magnitude intervals for the Field F4 with a curve of single mode 
indicating the thin disc. There are slight differences between the number 
of observed stars and the predicted ones. The arrows at the top each panel 
show the limiting completeness of $z$-distances for the bright and faint 
apparent magnitudes.} 
\end{center}
\end {figure}

\begin{figure}
\begin{center}
\includegraphics[angle=0, width=130.5mm, height=142.5mm]{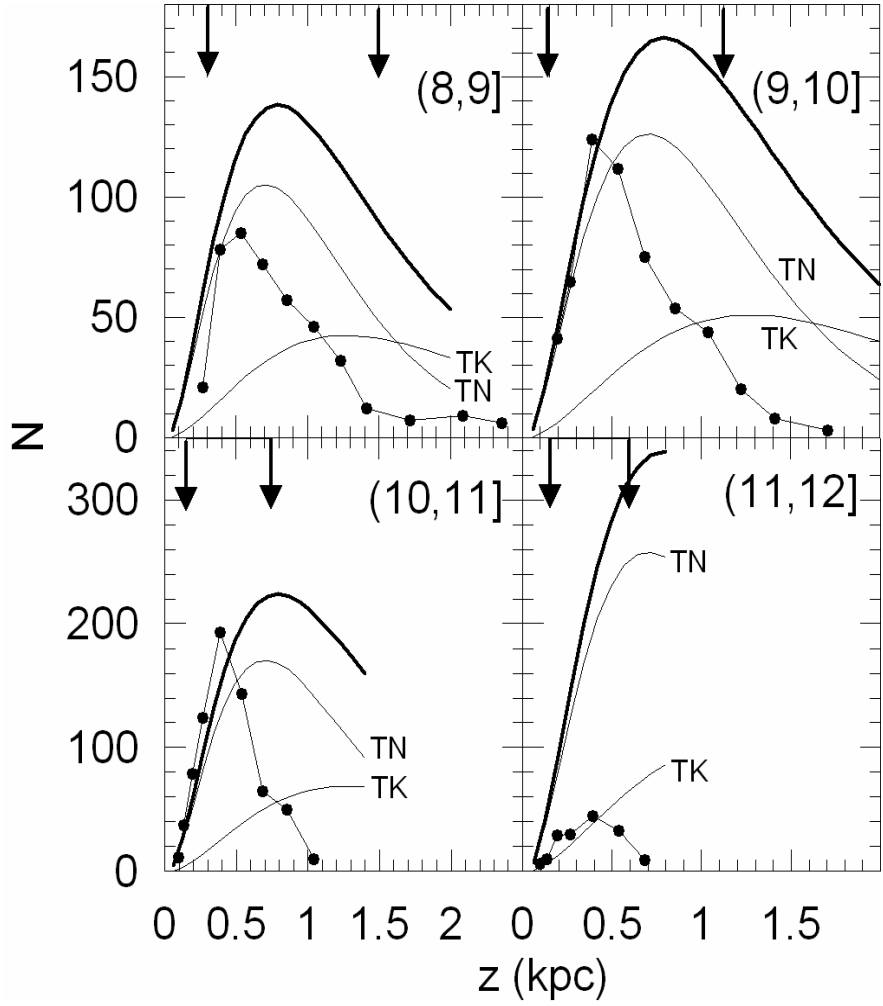}
\caption[] {Comparison of the distribution of distances above the Galactic plane 
for stars in the field F4, as an example, with the Galactic model of \cite{Chen01}. 
The scaleheight and solar normalization of the thick disc are $H=600$ pc, 
$n_{2}/n_{1}=12$ per cent, respectively. Small differences between the number of 
stars and the predicted ones for the thin disc (TN) indicate slight contamination of 
thick disc (TK). Arrows are as in the Fig. 5.} 
\end{center}
\end {figure}

\begin{figure}
\begin{center}
\includegraphics[angle=0, width=130.5mm, height=142.5mm]{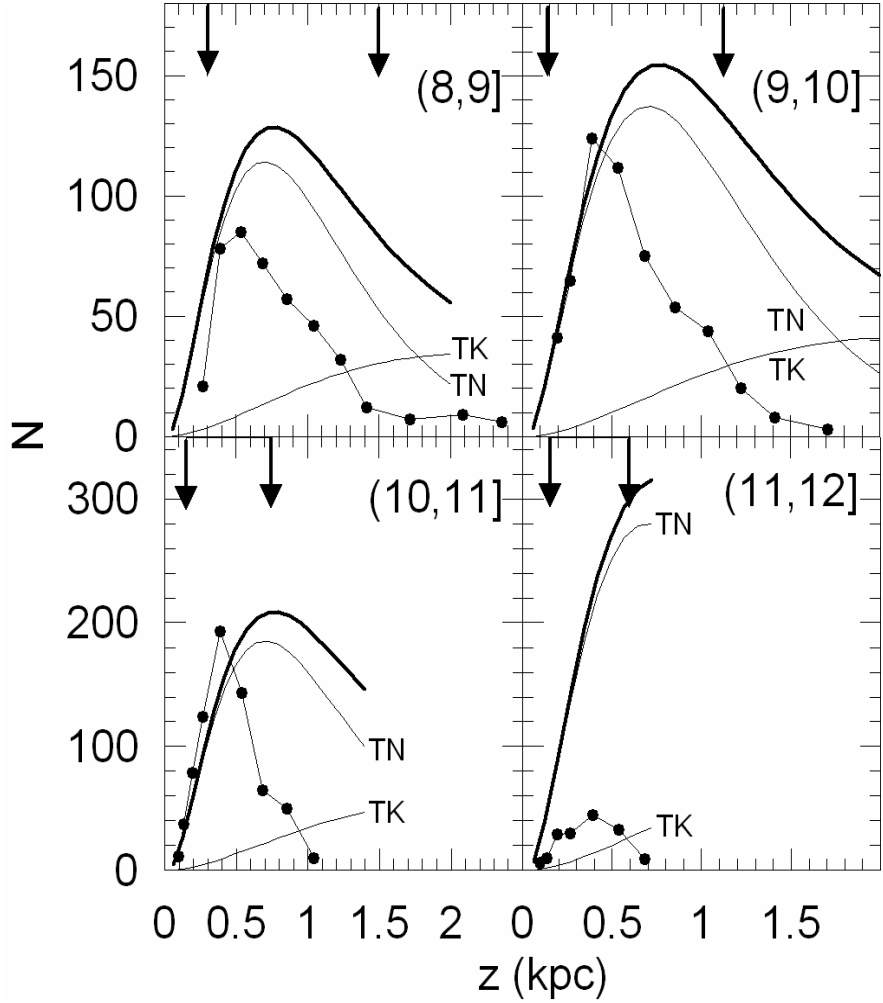}
\caption[] {Comparison of the distribution of distances above the Galactic plane 
for stars in the field F4, as an example, with the Galactic model of \cite{Chen01}. 
The scaleheight and solar normalization of the thick disc are $H=1000$ pc, 
$n_{2}/n_{1}=3$ per cent, respectively. The other model parameters are the same as
in Fig. 6. The same small discrepancy in Fig. 6 can be seen also here. Arrows are as 
in the Fig. 5.} 
\end{center}
\end {figure}

\begin{figure}
\begin{center}
\includegraphics[angle=0, width=130.5mm, height=142.5mm]{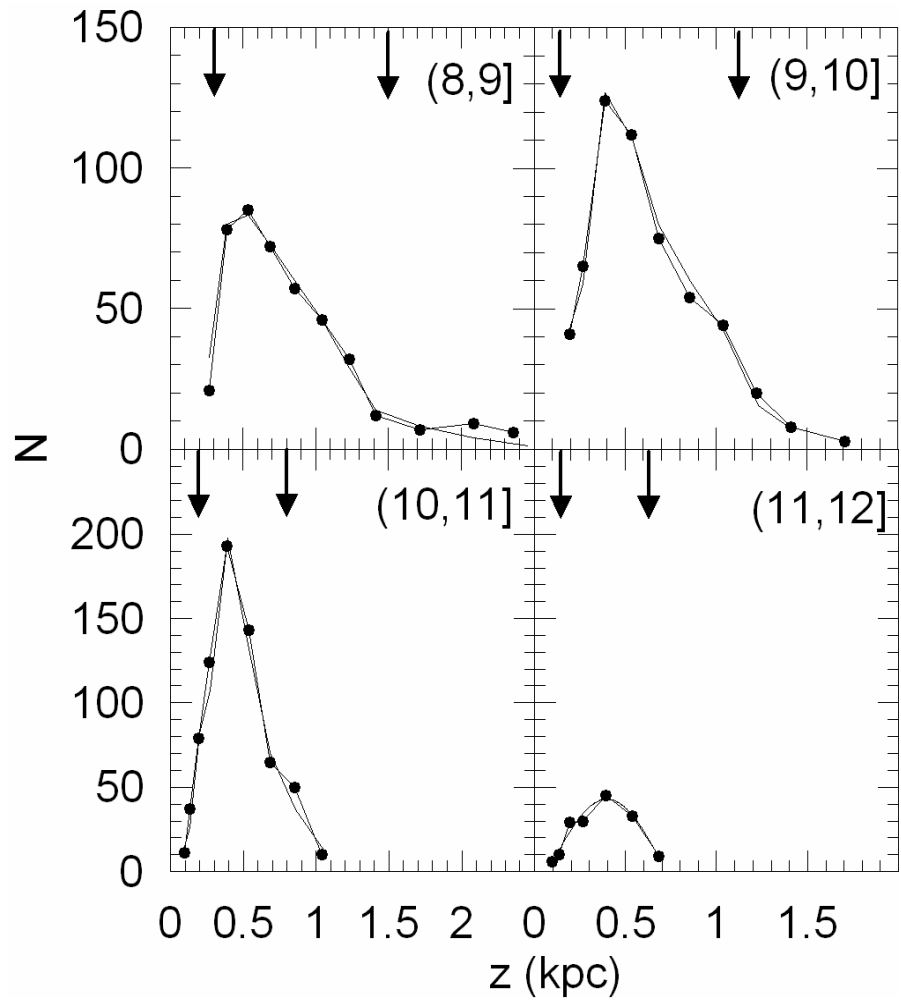}
\caption[] {Reduction of the number of stars by 7, 14, 16 per cent for the absolute 
magnitude intervals $8<M(g)\leq9$, $9<M(g)\leq10$, and $10<M(g)\leq11$ brings the 
observed number of stars close to the predicted ones. Arrows are as in the Fig. 5.}
\end{center}
\end {figure}

\begin{figure}
\begin{center}
\includegraphics[angle=0, width=121.5mm, height=180mm]{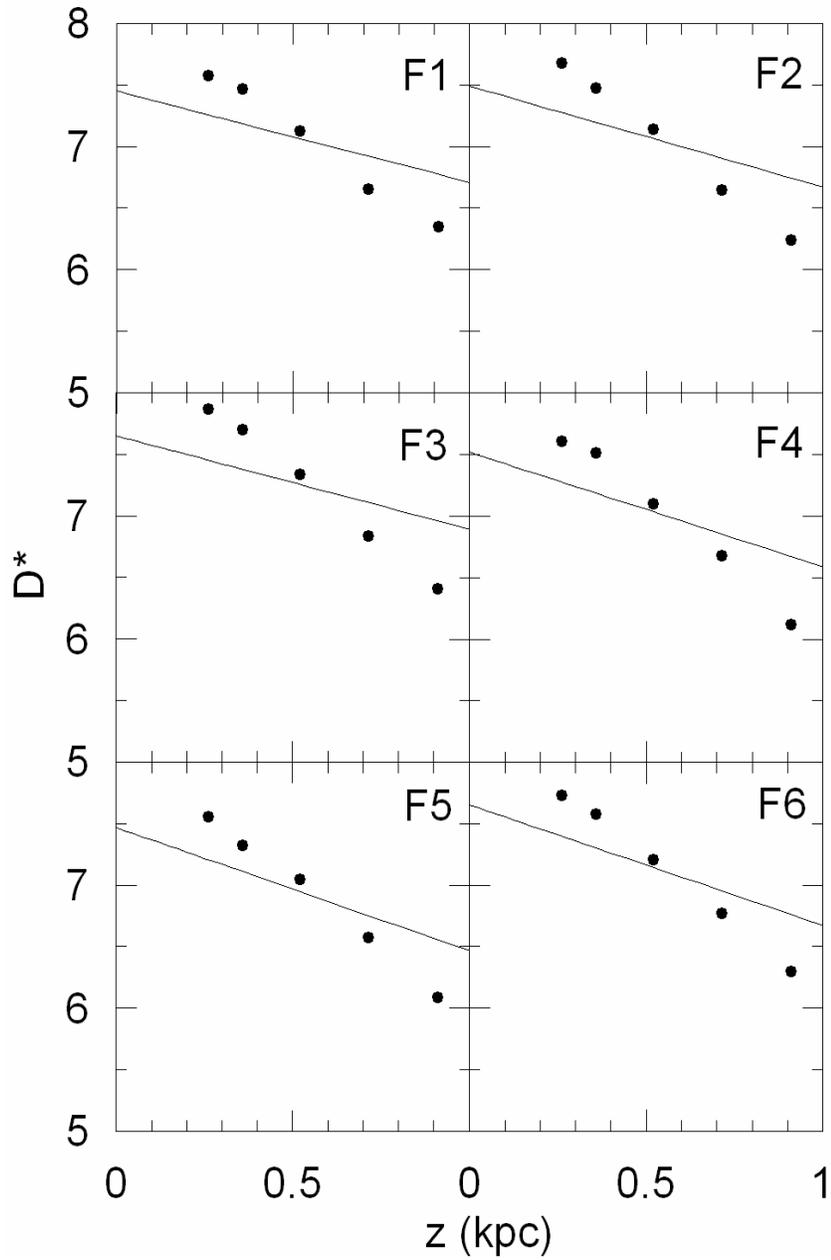}
\caption[] {Comparison of the derived logarithmic space densities $D^{*}$ 
with the Galactic disc model of \cite{Chen01} for stars with absolute 
magnitudes $10<M(g)\leq11$, for six fields. Agreement is rather poor and the 
standard deviations are large (see Table 5).}
\end{center}
\end {figure}

\begin{figure}
\begin{center}
\includegraphics[angle=0, width=150mm, height=171mm]{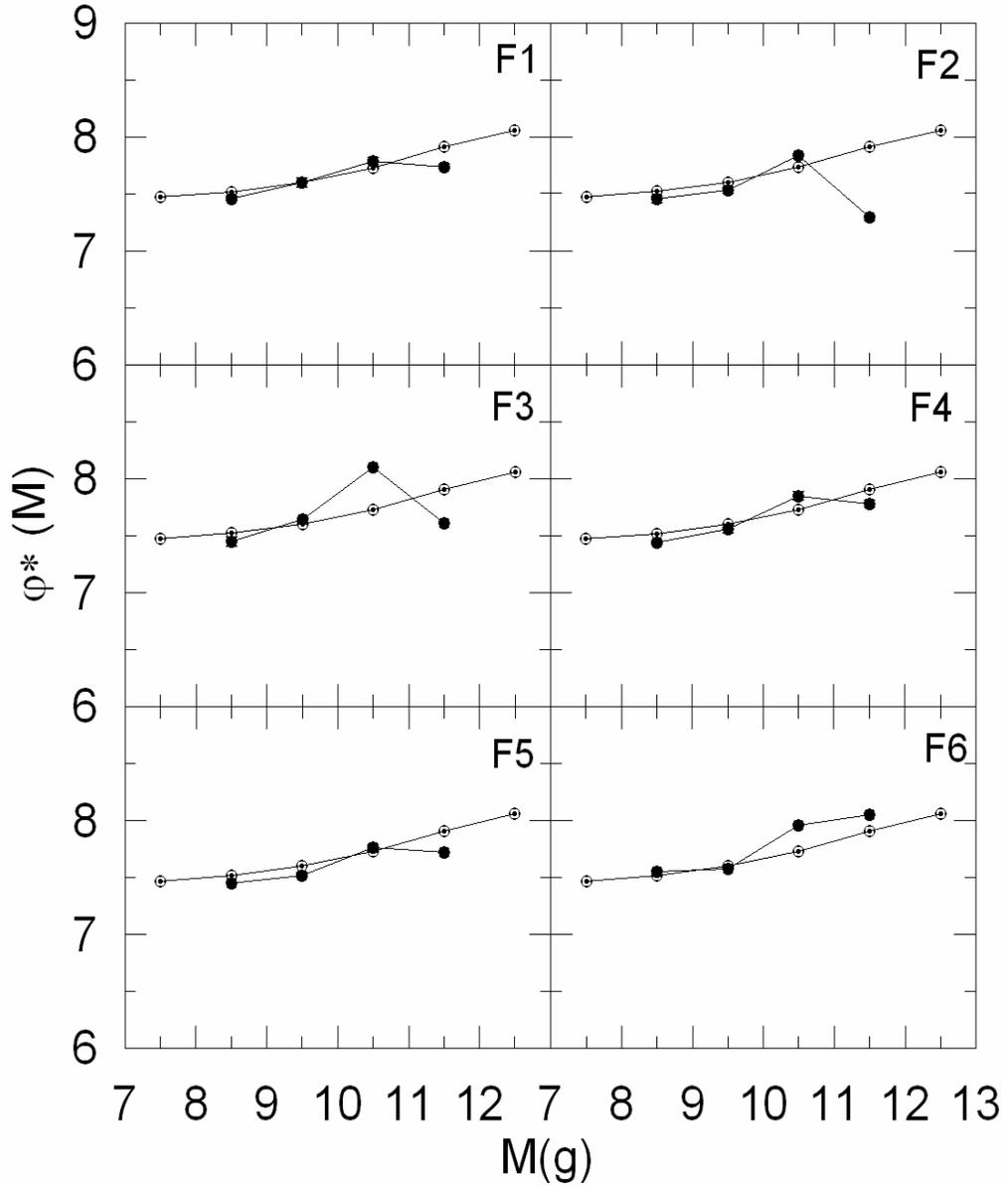}
\caption[] {The local luminosity function for the thin disc. The $\odot$ 
symbols indicate the Hipparcos standard values.}
\end{center}
\end {figure}

\begin{figure}
\begin{center}
\includegraphics[angle=0, width=102.5mm, height=180mm]{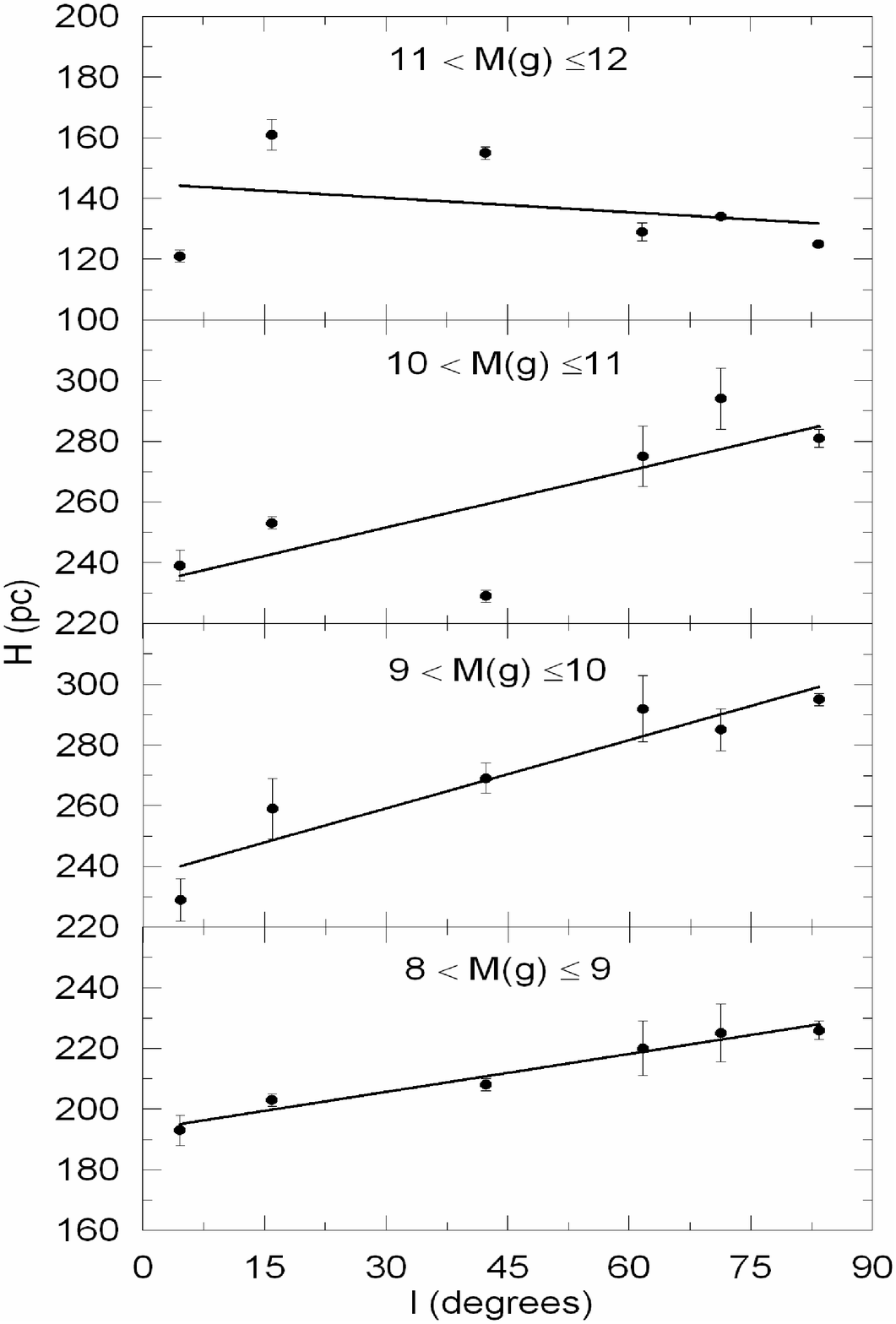}
\caption[] {The relation between the scaleheights for six fields and the 
corresponding Galactic longitudes.}
\end{center}
\end {figure}

\begin{figure}
\begin{center}
\includegraphics[angle=0, width=102.5mm, height=180mm]{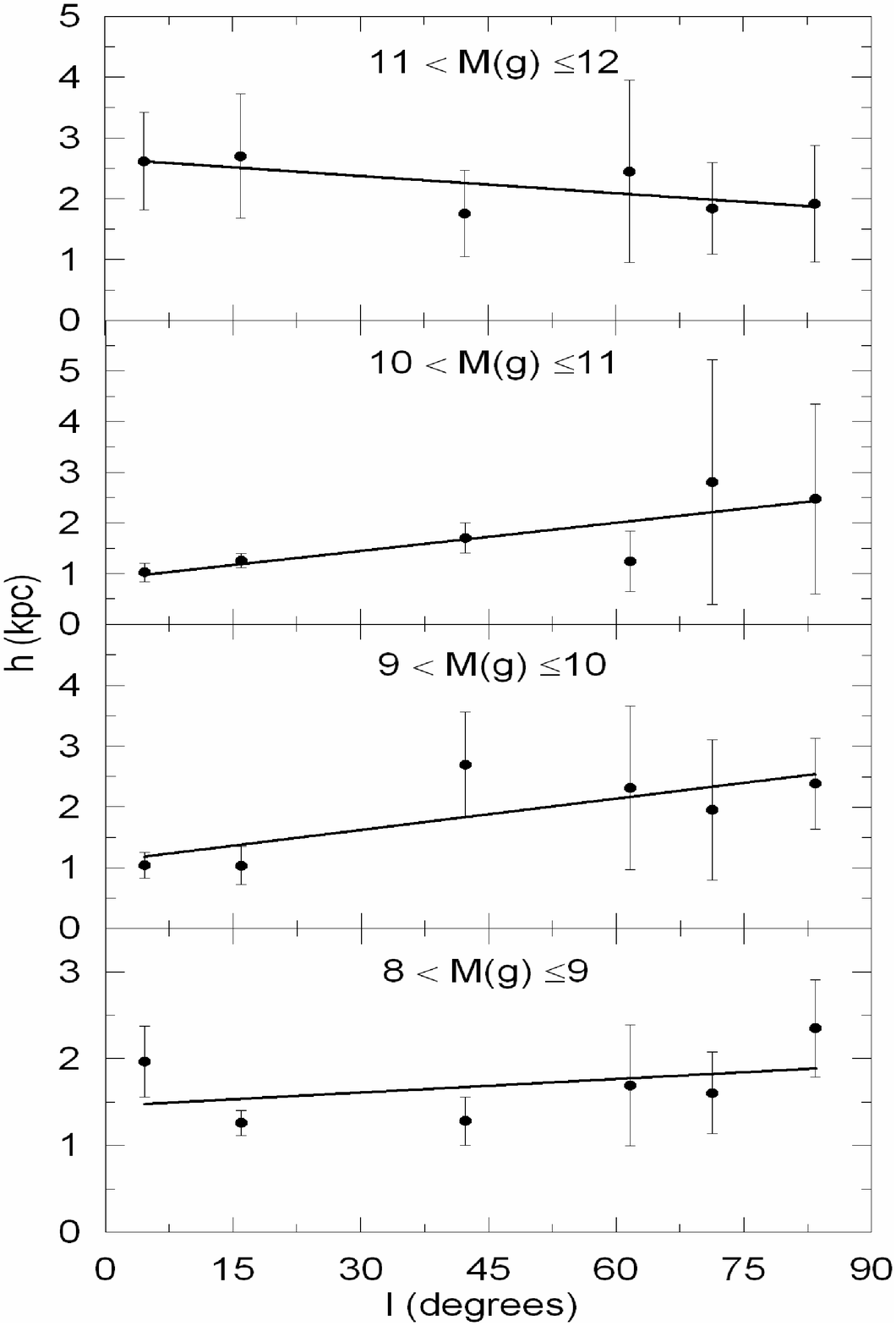}
\caption[] {The relation between the scalelengths for six fields and the 
corresponding Galactic longitudes.} 
\end{center}
\end {figure}

\begin{table}
\center
\caption{Previous Galactic models. Symbols: H: scaleheight, h: scalelenght, 
n: local space density of the population relative to the space density of the 
thin disc, $R_{e}$: the effective radius, and $\kappa=c/a$: the axis ratio.}
{\tiny
\begin{tabular}{lllllllll}
\hline
\multicolumn{2}{c}{Thin disc} & \multicolumn{3}{c}{Thick disc} & \multicolumn{3}{c}{Halo}&\\
H (pc)& h (kpc)& n$_{thick}$ & H (kpc)& h (kpc)& n$_{halo}$ & $R_{e}$ (kpc)& $\kappa$&  Reference \\
\hline
310-325 & --- & 0.0125-0.025 &  1.92-2.39 & --- & --- & --- & --- & \cite{Yoshii82}\\
300 & --- & 0.02 & 1.45 & --- & 0.0020 & 3.0 & 0.85 & \cite{GR83}\\
325 & --- & 0.02 & 1.3 &  --- & 0.0020 & 3.0 & 0.85 & \cite{Gilmore84}\\
280 & --- & 0.0028 &  1.9 & --- &0.0012 &--- & --- & \cite{TM84}\\
200-475 & --- & 0.016 &  1.18-2.21 & --- & 0.0016 & --- & 0.80 & \cite{RC86}\\
300 & --- & 0.02 & 1.0 & --- & 0.0010 & --- & 0.85 & \cite{DF87}\\
285 & --- & 0.015 &1.3-1.5 & --- & 0.0020 & 2.36 & flat & \cite{F87}\\
325 & --- & 0.0224 & 0.95 & --- & 0.0010 & 2.9 & 0.90 & \cite{Yoshii87}\\
249 & --- & 0.041 & 1.0 & --- & 0.0020 & 3.0 & 0.85 & \cite{KG89}\\
350 & 3.8 & 0.019 & 0.9 & 3.8 & 0.0011 & 2.7 & 0.84 & \cite{YY92}\\
290 & --- & --- & 0.86 & --- & --- & 4.0 & --- & \cite{HB93}\\
325 & --- & 0.0225 & 1.5 & --- & 0.0015 & 3.5 & 0.80 & \cite{RW93}\\
325 & 3.2 & 0.019 & 0.98 & 4.3 & 0.0024 & 3.3 & 0.48 & \cite{L96}\\
250-270 & 2.5 & 0.056 & 0.76 & 2.8 & 0.0015 & 2.44-2.75$^{a}$ &  0.60-0.85 & \cite{R96, R00}\\
290 &4.0 & 0.059 & 0.91 & 3.0 & 0.0005 & 2.69 & 0.84 & \cite{Buser98, Buser99}\\
240 &2.5 & 0.061 & 0.79 & 2.8 & --- & --- & 0.60-0.85 & \cite{O99}\\
330 &2.25& 0.065-0.13 &  0.58-0.75 & 3.5 & 0.0013 & --- & 0.55 & \cite{Chen01}\\
280 & 2-2.5&  0.06-0.10 & 0.7-1.0 & 3-4 & 0.0015 & --- & 0.50-0.70 & \cite{Siegel02}\\
350 & 2-2.5&  0.06-0.10 & 0.9-1.2 & 3-4 & 0.0015 & --- & 0.50-0.70 & \cite{Siegel02}$^{b}$\\
320 & --- & 0.07 & 0.64 & --- & 0.0013 & --- & 0.58 & \cite{Du03} \\
\hline
\multicolumn{4}{l}{a Power-law index replacing $R_{e}$.}\\
\multicolumn{8}{l}{b Corrected values for binarism.}
\end{tabular}
}
\end{table}

\begin{table}
\center
\caption{Data for six fields investigated in this work. The coordinates 
are for the epoch 2000, and N is the number of stars.}
\begin{tabular}{rccccrcc}
\hline
Field &  $\alpha$ &  $\delta$ & {\it l} & {\it b} & Size & $E(B-V)$ &        N \\
      & (h~~m~~s)&  ($^{o}~~^{'}~~^{''}$)& ($^{\circ}$)& ($^{\circ}$) & (deg$^2$)&   &    \\
\hline
   F1 &  15 34 14 &  -00 22 44 &    4.58 &   42.19 &   10 &    0.102 &  20 549 \\
   F2 &  15 40 00 &   08 30 00 &   15.93 &   46.06 &   10 &    0.044 &  19 198 \\
   F3 &  16 30 00 &   24 06 00 &   42.28 &   41.02 &   10 &    0.059 &  25 577 \\
   F4 &  16 02 22 &   38 38 19 &   61.62 &   48.78 &   10 &    0.013 &  14 523 \\
   F5 &  15 42 00 &   44 24 00 &   71.29 &   51.89 &   10 &    0.019 &   8 731 \\
   F6 &  09 52 00 &   52 45 00 &   83.38 &   48.55 &   20 &    0.015 &  24 802 \\
\hline
\end{tabular}
\end{table}  

\begin{table}
\center
\caption{Logarithmic space density function, $D^{*}=\log D+10$, for different 
absolute magnitude intervals for the Field F4 as an example. Distances in kpc, 
volumes in pc$^3$, horizontal lines show the limiting distance of completeness. 
Other symbols are explained in the text.}
\small{
\begin{tabular}{lrrrrrrrrrrrr}
\hline
\multicolumn{3}{c}{$M(g)\rightarrow$} & \multicolumn{2}{c}{(8,9]}& \multicolumn{2}{c}{(9,10]} & \multicolumn{2}{c}{(10,11]} & \multicolumn{2}{c}{(11,12]} &\multicolumn{2}{c}{(12,13]}\\
$r_{1}-r_{2}$ &$\Delta V_{1,2}$ &$r^{*}$ & N & $D^{*}$ & N & $D^{*}$ & N & $D^{*}$ & N & $D^{*}$ & N & $D^{*}$\\
\hline
 0.10-0.15 &2.41 (3) &0.13 &      &      &       &       &  11 &  7.66 &  6 & 7.40 &   &      \\
 0.15-0.20 &4.70 (3) &0.18 &      &      &       &       &  37 &  7.90 & 10 & 7.33 &   &      \\ \cline{6-9}
 0.20-0.30 &1.93 (4) &0.26 &      &      &  41   & 7.33  &  79 &  7.61 & 29 & 7.18 & 1 & 5.71 \\ \cline{10-11}
 0.30-0.40 &3.76 (4) &0.36 & 21   & 6.75 &  65   & 7.24  & 124 &  7.52 & 30 & 6.90 &   &      \\ \cline{4-5}
 0.40-0.60 &1.54 (5) &0.52 & 78   & 6.70 & 124   & 6.90  & 193 &  7.10 & 45 & 6.46 &   &       \\
 0.60-0.80 &3.01 (5) &0.71 & 85   & 6.45 & 112   & 6.57  & 143 &  6.68 & 33 & 6.04 &   &       \\ \cline{10-11}
 0.80-1.00 &4.96 (5) &0.91 & 72   & 6.16 &  75   & 6.18  &  65 &  6.12 &  9 & 5.26 &   &       \\ \cline{8-9}
 1.00-1.25 &9.68 (5) &1.14 & 57   & 5.77 &  54   & 5.75  &  50 &  5.71 &    &      &   &       \\
 1.25-1.50 &1.44 (6) &1.39 & 46   & 5.50 &  44   & 5.48  &  10 &  4.84 &    &      &   &       \\ \cline{6-7}
 1.50-1.75 &2.01 (6) &1.64 & 32   & 5.20 &  20   & 5.00  &     &       &    &      &   &       \\
 1.75-2.00 &2.68 (6) &1.88 & 12   & 4.65 &   8   & 4.47  &     &       &    &      &   &       \\ \cline{4-5}
 2.00-2.50 &7.74 (6) &2.28 &  7   & 3.96 &   3   & 3.59  &     &       &    &      &   &       \\
 2.50-3.00 &1.16 (7) &2.77 &  9   & 3.89 &       &       &     &       &    &      &   &       \\
 3.00-3.50 &1.61 (7) &3.27 &  6   & 3.57 &       &       &     &       &    &      &   &       \\
& \multicolumn{2}{c}{Total}&425   &      & 546   &       & 712 &       &162 &      & 1 &       \\     
\hline
\end{tabular}
}  
\end{table}

\begin{table}
\center
\caption{Galactic model parameters for different absolute magnitude 
intervals for the thin disc of six fields, resulting from the fits of derived 
and analytical density profiles. The columns indicate: (1) absolute magnitude 
interval $M(g)$, (2) density law, (3) and (8) logarithmic local space density 
$n^{*}$, (4) and (9) standard deviation for logarithmic space density $s$, (5) 
and (10) scaleheight (in pc) reduced to the exponential law $H$, (6) and (11) 
scalelength (in pc) $h$, (7) and (12) $\chi^{2}_{min}$, and (13) the standard local 
space density of Hipparcos reduced to the $SDSS$ photometry $\odot$ (see 
Appendix B).}
{\tiny
\begin{tabular}{rrrrrrrrrrrrr}
\hline
(1)   & (2)   & (3)   & (4)    & (5)   & (6)  & (7) & (8)  & (9)  & (10)  & (11)   & (12) & (13) \\
\hline
M(g)   & law   & $n^{*}$   & $s$    & $H$   & $h$  & $\chi^2_{min}$ & $n^{*}$   & $s$    & $H$   & $h$    &  $\chi^2_{min}$ & $\odot$ \\
\hline
&&\multicolumn{5}{c}{\bf Field 1}&\multicolumn{5}{c}{\bf Field 4}&\\
\hline
(8,9]&   exp&$7.46^{+0.02}_{-0.02}$&0.05& $193^{+4}_{-4}$&$1967^{+490}_{-330}$& 3.73& $7.44^{+0.02}_{-0.02}$&0.06& $220^{+4}_{-4}$& $1690^{+940}_{-450}$ & 3.34 &7.52\\
     &  sech& $7.17^{+0.02}_{-0.02}$& 0.05& $322^{+7}_{-7}$& $2169^{+580}_{-380}$ &3.82& $7.16^{+0.02}_{-0.02}$ & 0.06 & $360^{+7}_{-7}$ & $1675^{+970}_{-450}$ & 3.35&\\
     & sech$^{2}$& $6.98^{+0.03}_{-0.03}$ & 0.06 & $333^{+7}_{-7}$ &  $704^{+62}_{-50}$ & 6.26& $7.01^{+0.03}_{-0.03}$ & 0.06 & $427^{+9}_{-10}$ & $1158^{+620}_{-240}$ & 3.80 &\\
(9,10]&   exp & $7.90^{+0.04}_{-0.04}$ & 0.08 & $150^{+5}_{-5}$ & $2300^{+1750}_{-700}$ & 8.99 & $7.81^{+0.04}_{-0.04}$ & 0.06 & $183^{+9}_{-8} $& $2546^{+1880}_{-1560}$ & 6.05 &7.60\\
     &  sech & $7.60^{+0.04}_{-0.04}$ & 0.08 & $229^{+7}_{-7}$ & $1038^{+255}_{-170}$ & 8.98 & $7.56^{+0.03}_{-0.03}$ & 0.06 &$ 292^{+12}_{-10}$ & $2313^{+1450}_{-1250}$& 5.65&\\
     & sech$^{2}$ & $7.42^{+0.06}_{-0.04}$ & 0.10 & $277^{+12}_{-8}$ &  $935^{+200}_{-195}$& 14.74 & $7.48^{+0.02}_{-0.02}$ & 0.06 & $321^{+9}_{-9}$ & $1030^{+750}_{-300}$& 5.80&\\
(10,11]&   exp & $8.23^{+0.02}_{-0.02}$ & 0.06 & $129^{+2}_{-1}$ & $2417^{+1120}_{-580}$ &7.13& $8.21^{+0.06}_{-0.06}$ & 0.08 & $148^{+11}_{-9}$& $1589^{+1200}_{-1010}$ & 12.43&7.73 \\
      &  sech & $7.94^{+0.02}_{-0.02}$ & 0.05 & $197^{+3}_{-2} $&$ 1039^{+165}_{-130}$ & 4.21& $7.94^{+0.05}_{-0.04}$ & 0.08 &$ 242^{+15}_{-13}$ & $1867^{+1410}_{-1175}$ & 11.11&\\
      & sech$^{2}$ & $7.79^{+0.03}_{-0.03}$ & 0.04 &$ 239^{+5}_{-5} $&$ 1022^{+220}_{-150}$ & 4.04& $7.85^{+0.04}_{-0.04}$ & 0.04 & $275^{+10}_{-10}$ & $1237^{+650}_{-545}$ & 4.70&\\
(11,12] &   exp & $7.74^{+0.02}_{-0.02}$ & 0.03 & $121^{+2}_{-2}$ & $2619^{+1050}_{-540}$ & 0.66& $7.78^{+0.03}_{-0.03}$ & 0.03 & $129^{+3}_{-3}$ & $2448^{+1600}_{-1400}$ & 0.37 &7.91 \\
      &  sech & $7.45^{+0.02}_{-0.02}$ & 0.03 & $192^{+3}_{-3}$ & $1516^{+440}_{-280}$ &0.68 & $7.50^{+0.03}_{-0.03}$ & 0.03 & $209^{+5}_{-5}$ & $1730^{+740}_{-600}$ & 0.45& \\
      & sech$^{2}$& $7.30^{+0.03}_{-0.03}$ & 0.04 & $232^{+7}_{-5}$ & $1454^{+580}_{-325}$ & 1.19& $7.35^{+0.03}_{-0.03}$ & 0.04 & $252^{+5}_{-5}$ & $1580^{+880}_{-630}$ & 0.80\\
\hline
&&\multicolumn{5}{c}{\bf Field 2}&\multicolumn{5}{c}{\bf Field 5}&\\
\hline
(8,9] &   exp & $7.45^{+0.02}_{-0.03}$ & 0.03 & $203^{+3}_{-3}$ & $1260^{+170}_{-120}$ & 0.84 & $7.45^{+0.01}_{-0.01}$ & 0.03 & $225^{+2}_{-2}$ & $1606^{+600}_{-345}$ &0.59 &7.52 \\
      &  sech & $7.08^{+0.01}_{-0.01}$ & 0.04 & $255^{+3}_{-3}$ & $1326^{+80}_{-75}$ & 3.73& $7.15^{+0.01}_{-0.01}$ & 0.03 & $373^{+3}_{-3}$ & $1625^{+645}_{-360}$ & 0.59&\\
      & sech$^{2}$ & $6.87^{+0.01}_{-0.01}$ & 0.04 & $408^{+3}_{-3}$ &  $808^{+34}_{-32}$ &1.68 & $6.92^{+0.01}_{-0.01}$ & 0.03 & $465^{+3}_{-3}$ & $1030^{+155}_{-125}$ & 0.65&\\
(9,10] &   exp & $7.81^{+0.04}_{-0.04}$ & 0.06 & $162^{+6}_{-6}$ & $1135^{+450}_{-250}$ & 10.31 & $7.76^{+0.01}_{-0.01}$ & 0.04 & $180^{+3}_{-3}$ & $2080^{+1180}_{-980}$ & 2.50 & 7.60 \\
      &  sech & $7.53^{+0.05}_{-0.03}$ & 0.06 & $259^{+13}_{-7}$ & $1030^{+350}_{-275}$ & 10.06& $7.52^{+0.02}_{-0.02}$ & 0.04 & $285^{+7}_{-7}$ & $1950^{+1260}_{-1050}$ & 1.54&\\
      & sech$^{2}$ & $7.41^{+0.05}_{-0.05}$ & 0.08 & $312^{+14}_{-13}$ & $1051^{+455}_{-245}$ & 13.65& $7.44^{+0.05}_{-0.05}$ & 0.07 & $333^{+14}_{-17}$ & $2100^{+1710}_{-1425}$ & 8.70& \\
(10,11] &   exp & $8.28^{+0.02}_{-0.02}$ & 0.02 & $124^{+2}_{-2}$ & $1024^{+530}_{-500}$ & 1.30& $8.11^{+0.02}_{-0.02}$ & 0.04 & $155^{+8}_{-8}$ & $1114^{+640}_{-530}$ & 4.37&7.73 \\
      &  sech & $7.99^{+0.02}_{-0.02}$ & 0.02 & $217^{+3}_{-3}$ & $2150^{+630}_{-400}$ & 1.47& $7.85^{+0.03}_{-0.03}$ & 0.04 & $254^{+5}_{-5}$ & $2212^{+1720}_{-1430}$ & 3.34&\\
      & sech$^{2}$ & $7.84^{+0.01}_{-0.01}$ & 0.02 & $253^{+2}_{-2}$ & $1250^{+165}_{-120}$ & 1.24& $7.76^{+0.03}_{-0.03}$ & 0.03 & $294^{+11}_{-10}$& $2804^{+2450}_{-2030}$ & 1.68&\\
(11,12] & exp & $7.29^{+0.02}_{-0.02}$ & 0.02 & $161^{+5}_{-5}$ & $2703^{+1120}_{-930}$ & 0.17& $7.72^{+0.01}_{-0.01}$ & 0.02 & $134^{+1}_{-1}$ & $1845^{+825}_{-685}$ & 0.15&7.91 \\
      &  sech & $7.04^{+0.03}_{-0.03}$ & 0.03 & $234^{+8}_{-7}$ & $1010^{+310}_{-210}$ & 0.38& $7.44^{+0.01}_{-0.01}$ & 0.02 & $219^{+2}_{-2}$ & $1826^{+750}_{-625}$ & 0.17&\\
      & sech$^{2}$ & $6.97^{+0.04}_{-0.04}$ & 0.05 & $281^{+14}_{-13}$ & $1841^{+940}_{-700}$ & 0.78& $7.29^{+0.01}_{-0.01}$ & 0.02 & $264^{+1}_{-1}$ & $1604^{+320}_{-230}$ & 0.18&\\
\hline
&&\multicolumn{5}{c}{\bf Field 3}&\multicolumn{5}{c}{\bf Field 6}&\\
\hline
(8,9] &   exp & $7.45^{+0.04}_{-0.04}$ & 0.05 & $208^{+6}_{-6}$ & $1281^{+340}_{-220}$ & 3.11& $7.55^{+0.02}_{-0.02}$ & 0.07 & $226^{+4}_{-3}$ & $2350^{+1915}_{-1595}$ & 6.54 & 7.52 \\
      &  sech & $7.17^{+0.04}_{-0.04}$ & 0.05 & $368^{+12}_{-10}$ & $2979^{+1145}_{-955}$ & 6.39& $7.26^{+0.02}_{-0.02}$ & 0.07 & $371^{+7}_{-5}$ & $2266^{+1950}_{-1620}$ & 7.44\\
      & sech$^{2}$ & $6.99^{+0.03}_{-0.03}$ & 0.06 & $442^{+10}_{-10}$ & $1975^{+610}_{-380}$ & 7.30& $7.05^{+0.02}_{-0.02}$ & 0.09 & $469^{+7}_{-7}$ & $1010^{+620}_{-500}$ & 10.59&\\
(9,10] &   exp & $7.91^{+0.02}_{-0.02}$ & 0.06 & $159^{+4}_{-3}$ & $1520^{+510}_{-310}$ & 7.97& $7.87^{+0.01}_{-0.01}$ & 0.01 & $178^{+1}_{-1}$ & $1125^{+270}_{-190}$ & 0.12&7.60 \\
      &  sech & $7.64^{+0.02}_{-0.02}$ & 0.06 & $269^{+5}_{-5}$ & $2695^{+980}_{-755}$ & 7.42& $7.58^{+0.01}_{-0.01}$ & 0.01 & $295^{+2}_{-2}$ & $2382^{+800}_{-695}$ & 0.11&\\
      & sech$^{2}$ & $7.53^{+0.02}_{-0.02}$ & 0.07 & $307^{+7}_{-7}$ & $1680^{+580}_{-345}$ &  10.72& $7.42^{+0.01}_{-0.01}$ & 0.02 & $358^{+3}_{-3}$ & $1370^{+705}_{-590}$ & 1.06&\\
(10,11] &   exp & $8.54^{+0.01}_{-0.01}$ & 0.02 & $116^{+1}_{-1}$ & $2138^{+560}_{-390}$ &  1.28& $8.30^{+0.02}_{-0.02}$ & 0.04 & $152^{+4}_{-4}$ & $1065^{+890}_{-745}$ & 7.94&7.73 \\
      &  sech & $8.25^{+0.01}_{-0.01}$ & 0.02 & $192^{+2}_{-2}$ & $2447^{+750}_{-450}$ & 2.23 & $8.04^{+0.02}_{-0.02}$ & 0.03 & $245^{+7}_{-5}$ & $1820^{+1680}_{-1400}$ & 4.24&\\
      & sech$^{2}$ & $8.10^{+0.01}_{-0.01}$ & 0.02 & $229^{+2}_{-2}$ & $1704
^{+350}_{-240}$ & 1.23& $7.96^{+0.01}_{-0.01}$ & 0.01 & $281^{+3}_{-3}$ & $2470^{+2040}_{-1700}$ & 1.30&\\
(11,12] &   exp & $7.61^{+0.01}_{-0.01}$ & 0.01 & $155^{+2}_{-2}$ & $1755^{+940}_{-480}$ & 0.01& $8.05^{+0.01}_{-0.01}$ & 0.01 & $125^{+1}_{-1}$ & $1920^{+1050}_{-875}$ & 0.13 & 7.91 \\
      &  sech & $7.39^{+0.01}_{-0.01}$ & 0.01 & $242^{+3}_{-3}$ & $2720^{+565}_{-405}$ & 0.03& $7.76^{+0.01}_{-0.01}$ & 0.01 & $206^{+2}_{-2}$ & $1295^{+660}_{-550}$ & 0.17&\\
      & sech$^{2}$ & $7.34^{+0.01}_{-0.01}$ & 0.01 & $264^{+1}_{-1}$ & $2985^{+155}_{-100}$ & 0.10& $7.63^{+0.01}_{-0.01}$ & 0.03 & $245^{+3}_{-3}$ & $1800^{+1445}_{-1205}$ & 0.94&\\
\hline
\end{tabular}
}
\end{table}

\begin{table}
\center
\small{
\caption{Logarithmic solar space densities, $n^{*}$, for stars with absolute magnitudes 
$10<M(g)\leq 11$ for six fields estimated by fitting the derived space densities with the 
Galactic disc model of \cite{Chen01}. The symbol $s$ is the standard deviation. The local 
logarithmic solar space density of Hipparcos \citep{Jahreiss97} is $\odot=7.73$.} 
\begin{tabular}{ccc}
\hline
Field &    $ n^{*}$ &  $s$\\
\hline
F1 & 7.45 & 0.33\\
F2 & 7.49 & 0.37\\
F3 & 7.65 & 0.40\\
F4 & 7.52 & 0.37\\
F5 & 7.47 & 0.33\\
F6 & 7.65 & 0.33\\
\hline
\end{tabular}
}  
\end{table}

\begin{table}
\center
\small{
\caption{The scaleheight, $H$ (pc), and scalelength, $h$ (pc), as a function of both 
absolute magnitude and longitude for six fields.}
\begin{tabular}{cccccccccccccc}
\hline
\multicolumn{2}{c}{$M(g)\rightarrow$} & \multicolumn{3}{c}{(8,9]} &  \multicolumn{3}{c}{(9,10]}& 
\multicolumn{3}{c}{(10,11]}& \multicolumn{3}{c}{(11,12]} \\
\hline
Field & $l(^{o})$ & n$^{*}$ & $H$ & $h$ & n$^{*}$ & $H$ & $h$ & n$^{*}$ & $H$ & $h$ & n$^{*}$ & $H$ & $h$\\
\hline
F1 &  4.58 & 7.46 & 193 & 1967 & 7.60 & 229 & 1038 & 7.79 & 239 & 1022 & 7.74 & 121 & 2619 \\
F2 & 15.93 & 7.45 & 203 & 1260 & 7.53 & 259 & 1030 & 7.84 & 253 & 1250 & 7.29 & 161 & 2703 \\
F3 & 42.28 & 7.45 & 208 & 1281 & 7.64 & 269 & 2695 & 8.10 & 229 & 1704 & 7.61 & 155 & 1755 \\
F4 & 61.62 & 7.44 & 220 & 1690 & 7.56 & 292 & 2313 & 7.85 & 275 & 1237 & 7.78 & 129 & 2448 \\
F5 & 71.29 & 7.45 & 225 & 1606 & 7.52 & 285 & 1950 & 7.76 & 294 & 2804 & 7.72 & 134 & 1845 \\
F6 & 83.38 & 7.55 & 226 & 2350 & 7.58 & 295 & 2382 & 7.96 & 281 & 2470 & 8.05 & 125 & 1920 \\
\hline
\multicolumn{3}{c}{Average}& 212 & 1692 &  & 272 & 1901 & &   262 & 1748 &  &  138 & 2215 \\
\hline
\end{tabular} 
} 
\end{table}

\begin{table}
\center
\caption{Coefficients $a_{i}$ and $b_{i}$ (i=0 and 1) in the eqs. (8) and (9) and the 
squared correlation coefficient R$^{2}$.}
\begin{tabular}{ccccccc}
\hline
  &\multicolumn{3} {c} {$H=a_{1}(l)+a_{0}$}  & \multicolumn{3} {c} {$h=b_{1}(l)+b_{0}$}  \\
\hline
      M(g) &    $a_{1}$ &     $a_{0}$&    $R^{2}$ &    $b_{1}$ &    $b_{0}$ &    $R^{2}$ \\
\hline
    (8, 9] &      0.417 &      193.1 &       0.96 &       0.01 &      1.451 &       0.15 \\
   (9, 10] &      0.750 &      236.6 &       0.89 &       0.02 &      1.100 &       0.57 \\
  (10, 11] &      0.625 &      232.8 &       0.59 &       0.02 &      0.887 &       0.63 \\
  (11, 12] &     -0.158 &      144.9 &       0.09 &      -0.01 &      2.654 &       0.49 \\
\hline
\end{tabular}  
\end{table}
\end{document}